\documentstyle[preprint,prl,aps,psfig]{revtex}
\begin{document}

\draft

\title{Drag in Paired Electron-Hole Layers}
\author{Giovanni Vignale}
\address{Department of Physics, University of Missouri, Columbia, MO,
65211, USA}
\author{A.H. MacDonald}
\address{Department of Physics, Indiana University, Bloomington,
IN 47405, USA}

\date{\today}
\maketitle

\begin{abstract}
\leftskip 2cm
\rightskip 2cm

We investigate transresistance effects in electron-hole double-layer
systems with an excitonic condensate.  Our theory is based on the use of
a minimum dissipation premise to fix the current carried by the
condensate.  We find that the drag resistance jumps discontinuously
at the condensation temperature and diverges as the temperature
approaches zero.

\end{abstract}



The possibility of realizing an excitonic condensate
in a system composed of spatially separated layers of electrons and holes
was proposed some time ago.\cite{oldrussian}  Only recently, however
has it become feasible\cite{fukuzawa,kash,sivan,kane} to produce systems where
 the
electrons and holes are close enough together to interact strongly
and at the same time sufficiently isolated to strongly inhibit optical
recombination.  Hopes that an excitonic condensate might
occur are supported by theoretical work in the strong magnetic
field limit\cite{sfcond,dlcoh} where some simplifications occur
and the conclusion can be established with greater confidence.
In this Letter, we present a theory of the experimental signature
of an excitonic condensate in the transport properties of a
coupled electron-hole double-layer (EHDL) system.  Interest in the coupling
of transport coefficients of nearby electron and hole layers due to
interactions, long recognized as a theoretical\cite{dragold} possibility,
has increased lately\cite{recenttheory} with the advent of
accurate experimental measurements.\cite{sivan,dragexpts}  The coupling
is revealed most starkly in measurements of the transresistivity;
the ratio of the electric field in an electrically
open layer to the current density flowing in a nearby layer.
The transresistivity is typically\cite{dragexpts} several
orders of magnitude smaller than the isolated layer resistivity.
In our theory, the transresistivity jumps to a value comparable
to the isolated layer resistivity as soon as the condensate forms,
it continues to increase with decreasing temperature $T$, and
it diverges as $T \to 0$.

The starting point for our theory is a minimum dissipation\cite{whodowecite}
premise which we use to fix the pair momentum of the
condensed excitons in the non-equilibrium current carrying state
of the EHDL system.  Using a matrix notation for the layer indices ($e$ for
electrons, $h$ for holes) we partition the total current density into a
superfluid portion (${\bf j_s} = (j_{sh},j_{se})$) carried by the condensate
and
a normal portion (${\bf j_n}=  (j_{nh},j_{ne})$) carried by the quasiparticles:
\begin{equation}
{\bf j} = {\bf j_s} +{\bf j_n}
\label{eq:current}
\end{equation}
where the superfluid portion is proportional to the pairing
momentum $P$.
($\vec P$ will be directed along the direction of current flow.)
Since it represents the flow of oppositely charged
bound pairs ${\bf j_s} = (P e n_{pair}/ m_{pair}) (1,-1)$ where
$n_{pair}/m_{pair}$ is defined by this equation.  The physical situations
of interest to us will include ones in which the sum of electron and hole
currents is not zero so that ${\bf j_n}$ cannot
be zero.  A quasiparticle current will be generated if electric fields
which drive the quasiparticles from equilibrium with the condensate are
present in the electron and hole layers.  In linear response
\begin{equation}
{\bf j_n} = {\bf \sigma^{qp}}{\bf E}
\label{eq:normcurrent}
\end{equation}
(Expressions for $n_{pair}/m_{pair}$ and the quasiparticle
transconductivity matrix $\bf \sigma^{qp}$ based on microscopic theory will
be derived below.)  A quasiparticle current flowing in the presence
of electric fields will dissipate energy at a rate per unit area given
by the Joule heating expression:
\begin{equation}
W(P) = {\bf j_n}\cdot{\bf E}=({\bf j} - Pe n_{pair}/m_{pair}(1,-1))
\cdot {\bf \rho^{qp}} \cdot ({\bf j} - Pe n_{pair}/m_{pair}(1,-1))
\label{eq:dissipation}
\end{equation}
where ${\bf \rho^{qp}}$, the quasiparticle transresistivity matrix, is the
 inverse
of the quasiparticle transconductivity matrix.
The pairing momentum of the condensate
in the non-equilibrium current carrying state will be the one which
allows the prescribed currents to flow through
the system with minimum dissipation.  Applying this condition we find
a surprisingly simple result;
\begin{equation}
{\bf j_{s}}\cdot{\bf E} =0
\label{eq:simple}
\end{equation}
so that {\it the electric fields in electron and hole layers are identical}
independent of the microscopic transresistivity matrix and the
prescribed currents.

Explicit expressions for ${\bf j_s}$, ${\bf j_n}$ and ${\bf E}$
can be obtained by combining Eq.(~\ref{eq:current}),
Eq.(~\ref{eq:normcurrent}), and Eq.(~\ref{eq:simple}).  We find
that for prescribed current ${\bf j}=(j_h,j_e)$,
\begin{equation}
j_{sh}=-j_{se} = \frac{(\rho^{qp}_{hh}-\rho^{qp}_{eh})j_h
-(\rho^{qp}_{ee}-\rho^{qp}_{eh})j_e}
{\rho^{qp}_{ee}+\rho^{qp}_{hh}-2 \rho^{qp}_{eh}},
\label{eq:supercurrent}
\end{equation}
${\bf j_n}={\bf j} - {\bf j_s}$, and
\begin{equation}
E_h=E_e=\rho_{cd} (j_e+j_h).
\label{eq:drag}
\end{equation}
In Eq.(~\ref{eq:drag}) $\rho_{cd}$, which we will refer to as the
condensate drag resistivity, is related to the components of
the quasiparticle transresistivity and transconductivity matrices by:
\begin{eqnarray}
\rho_{cd} &=& \frac{\rho^{qp}_{ee} \rho_{hh}^{qp} - (\rho^{qp}_{eh})^2}
{\rho^{qp}_{ee}+\rho^{qp}_{hh}- 2 \rho^{qp}_{eh}} \nonumber\\
&=& \frac{1}{\sigma^{qp}_{ee}+\sigma_{hh}^{qp}+ 2 \sigma^{qp}_{eh}}.
\label{eq:rhocd}
\end{eqnarray}

We now turn our attention to the evaluation of $\rho_{cd}$ from
microscopic theory.  We will restrict our attention to the case
where the particle densities in electron and hole layers are
identical ($n_e = n_h =n$) and use a BCS mean-field theory\cite{tinkham}
to describe the pairing.  We will be able to express our results
in terms of the solution to the mean-field gap equations at
$P=0$.  Taking the effective attractive interaction $V$\cite{caveatbcs} which
enters the BCS equations to be independent of momentum, the gap equations
differ from their textbook counterparts only in that the
the electron and hole masses ($m_e$ and $m_h$) are not equal.
The quasiparticle energies are given by ($\hbar =1$)
\begin{eqnarray}
E_{0k} &=& E_k + \eta_k \nonumber\\
E_{1k} &=& E_k - \eta_k
\label{eq:qpengs}
\end{eqnarray}
where $E_k = (\epsilon_k^2 + \Delta^2)^{1/2}$, $\epsilon_k =
(k^2 -k_F^2)/ 2 m_{+}$, $\eta_k = (k^2 - k_F^2) / 2 m_{-}$,
$2 m_{\pm}^{-1}=m_e^{-1} \pm m_h^{-1}$, $k_F = (2 \pi n)^{1/2}$ is the Fermi
momentum, and $\Delta$ is determined by solving the gap equation:
\begin{equation}
\frac{1}{\lambda} = \int_0^{\omega_c} d \epsilon_k \frac{1}{\sqrt{\Delta^2 +
\epsilon_k^2}}[1 - f(E_{0k})-f(E_{1k})].
\label{eq:gap}
\end{equation}
In Eq.(~\ref{eq:gap}) $\lambda = N(0)V$  ($N(0)$ is the density of states
of free fermions of mass $m_+$ and density $n$) is the usual
dimensionless coupling constant\cite{tinkham} of BCS theory, $f(E) =
(\exp(E/k_B T)+1)^{-1}$, and $\omega_c$ is the cutoff for the
attractive interaction.

The microscopic calculations, which are somewhat lengthy, are similar to
common applications of BCS theory for superconductivity in metals.
The main steps of this calculation are sketched and the principal
results are given below.
We first calculate $n_{pair}/m_{pair}$ by evaluating the electron and hole
 currents
to first order in the pairing momentum $P$ when the quasiparticles are
in equilibrium with the condensate.  For $m_e$=$m_h$ this calculation
is identical to the calculation of the superfluid density which determines
the penetration depth of a superconductor.  We find that
\begin{equation}
\frac{n_{pair}}{m_{pair}} = \frac{n_e-n_{ne}}{m_e} = \frac{n_h-n_{nh}}{m_h}.
\label{eq:sfdens}
\end{equation}
The equivalence of the two forms for the right hand side of
Eq.(~\ref{eq:sfdens}) provides microscopic confirmation of the
expectation, used in our minimum dissipation analysis,
that the current carried by the electron-hole condensate is equal and
opposite in electron and hole layers.
In Eq.(~\ref{eq:sfdens}) the ``normal" densities $n_{ne}$ and $n_{nh}$ are
given
by

\begin{eqnarray}
n_{ne} &=& -\frac{1}{A} \sum_k \frac{k^{2}}{2m_{+}}\; \left[u_{k}^{2}
f'(E_{0k}) + v_{k}^{2} f' (E_{1k})\right]\nonumber\\
 && - \frac{1}{A} \sum_k \frac{k^{2}}{2m_{-}}\;
\frac{\epsilon_k}{E_{k}}\; \left[u_{k}^{2} f'
(E_{0k}) - v_{k}^{2} f' (E_{1k})\right]\nonumber\\
 && - \frac{1}{A}\sum_k \frac{k^{2}}{4 m_{-}}\;
\frac{\Delta^{2}}{E_{k}^{3}}\; \left( f (E_{0k}) + f
(E_{1k}) - 1\right),
\end{eqnarray}
where $A$ is the area of the layer, and
\begin{eqnarray}
\rho_{nh} &=& -\frac{1}{A}\sum_k \frac{k^{2}}{2m_{+}}\; \left[u_{k}^{2} f'
(E_{1k}) + v_{k}^{2} f' (E_{0k})\right]\nonumber\\
 && + \frac{1}{A}\sum_k \frac{k^{2}}{2m_{-}}\;
\frac{\epsilon_{k}}{E_{k}}\; \left[u_{k}^{2} f'
(E_{1k}) - v_{k}^{2} f' (E_{0k})\right]\nonumber\\
 && + \frac{1}{A}\sum_k \frac{k^{2}}{4m_{-}}\;
\frac{\Delta^{2}}{E_{k}^{3}}\; \left( f (E_{0k}) + f
(E_{1k}) - 1\right)
\end{eqnarray}
where $u_k^2 = (1+\epsilon_k/E_k)/2$ and $v_k^2= (1 -
\epsilon_k/E_k)/2$.

We have evaluated the quasiparticle transconductivity in the paired state
in a single-loop approximation using a Nambu-Gorkov Green's function
formalism\cite{mahan}.  After some standard manipulations this
approximation leads to
\begin{equation}
{\sigma^{qp}}_{ij} = \frac{1}{A} \sum_{\vec{p}} \frac{\pi p^{2}}{2 m_{i}
m_{j}} \int_{-\infty}^{\infty} f'(\omega) A_{ij}^{2}(\vec{p},\omega)\;
d\omega
\end{equation}
where $i$ and $j$ are layer indices,
\begin{equation}
{\bf A}(\vec{p},\omega) \equiv -\frac{1}{\pi} {\rm Im}\;
{\bf G}(\vec{p},\omega + i\delta)
\end{equation}
where ${\bf G}$ is the Nambu-Gorkov matrix Greens function.

In the absence of disorder
\begin{equation}
{\bf G}_{(0)}(\vec{p},\omega + i \delta) = \frac{(\omega - \eta_{p})
\hat{1} + \epsilon_{p} \hat{\tau}_{3} + \Delta
\hat{\tau}_{1}}{(\omega - E_{0p} + i\delta) (\omega + E_{1p} + i\delta)},
\end{equation}
where the $\tau$'s are Pauli matrices, so that
\begin{mathletters}
\begin{eqnarray}
A_{hh}^{(0)}(\vec{p},\omega) &=& u_{p}^{2} \delta (\omega - E_{0p}) +
v_{p}^{2} \delta (\omega + E_{1p})\\
A_{ee}^{(0)}(\vec{p},\omega) &=& v_{p}^{2} \delta (\omega - E_{0p}) +
u_{p}^{2} \delta (\omega + E_{1p})\\
A_{eh}^{(0)}(\vec{p},\omega) &=& \frac{\Delta}{2E_{p}}\; \left[\delta
(\omega - E_{0p}) - \delta (\omega + E_{1p})\right] =
A_{he}^{(0)}(\vec{p},\omega).
\end{eqnarray}
\end{mathletters}
and the quasiparticle conductivity diverges as expected.
To model disorder we
have included a self-consistent Born approximation\cite{mahan} self-energy
correction to the Nambu-Gorkov Green's function.  Assuming zero correlation
length for the disorder potential in each layer and no correlation between
the disorder in electron and hole layers a series of standard manipulations
allows the quasiparticle transconductivity to be expressed in terms of the
normal state scattering times for electron and hole layers
$\tau_{ne}$ and $\tau_{nh}$.  We find that
\begin{mathletters}
\begin{eqnarray}
\frac{\sigma^{qp}_{hh}}{\sigma_{0}} &=& \frac{(1 + y)^{2}}{2\alpha}
\int_{0}^{\infty} d\epsilon_k\;
\left[\frac{u_k^{4} \tilde{\tau}_{0k}}{\cosh^{2}(E_{0k}/2 k_BT)} +
\frac{v_k^{4} \tilde{\tau}_{1k}}{\cosh^{2}(E_{1k}/2k_BT)}\right]\\
\frac{\sigma^{qp}_{ee}}{\sigma_{0}} &=& \frac{(1 - y)^{2}}{2\alpha}
\int_{0}^{\infty} d\epsilon_k\;
\left[\frac{v_k^{4} \tilde{\tau}_{0k}}{\cosh^{2}(E_{k0}/2k_BT)} +
\frac{u_k^{4} \tilde{\tau}_{1k}}{\cosh^{2}(E_{1k}/2k_BT)}\right]\\
\frac{\sigma^{qp}_{eh}}{\sigma_{0}} &=& -\frac{1 - y^{2}}{2\alpha}
\int_{0}^{\infty} d\epsilon_k\; u_k^{2}v_k^{2}
\left[\frac{\tilde{\tau}_{0k}}{\cosh^{2}(E_{0k}/2k_BT)} +
\frac{\tilde{\tau}_{1k}}{\cosh^{2}(E_{1k}/2k_BT)}\right]
\end{eqnarray}
\end{mathletters}
where $y \equiv (m_e-m_h)/(m_h+m_e)$, $r \equiv (\tau_{ne}-\tau_{nh})/
(\tau_{nh}+\tau_{ne})$, $\alpha = (1+y)/(1+r)+(1-y)/(1-r)$,
$2 (\tau_n)^{-1} \equiv \tau_{nh}^{-1} + \tau_{ne}^{-1}$,
$\sigma_0 \equiv n e^2 \tau_n \alpha/ m_{+}$, and $\tilde {\tau}_{0k}$
and $\tilde {\tau}_{1k}$ are the energy dependent scattering times
which give the lifetimes of the quasiparticle states in units of $\tau_n$:
\begin{eqnarray}
\tilde{\tau}_{0k} &=& \frac{\epsilon_k/E_k +
y}{u_k^4(1+y)(1+r)+v_k^4(1-y)(1-r)}
\nonumber\\
\tilde{\tau}_{1k} &=& \frac{\epsilon_k/E_k
-y}{u_k^4(1-y)(1+r)+v_k^4(1+y)(1+r)}.
\label{eq:qplifetime}
\end{eqnarray}

In Fig.1 we show numerical results calculated from the
above expressions for the symmetric case $m_{e} = m_{h}$,
$\tau_{ne}=\tau_{nh}$. The results for the nonsymmetric case are
qualitatively similar.  The drag resistivity $\rho_{cd} \equiv
\sigma_{cd}^{-1}$ immediately jumps to a value comparable to the normal state
resistivity ($\sim \sigma_0^{-1}$) at $T_c$ and it diverges exponentially as
the
temperature goes to zero.  In this limit the quasiparticle conductivity
vanishes
because of the small number of thermally excited quasiparticles.
Consequently,  a larger and larger electric field is required to drive the
normal component of the current.  The behavior of the drag  conductivity for
$T$
just below $T_c$ can be obtained in analytic form:  in the
symmetric case $\sigma_{cd}/\sigma_0 \sim 1 - 2.49 (1 - T/T_c)^{1/2}$. We
also  plot the behavior of the quasiparticle conductivities
$\sigma^{qp}_{ee}$ and $\sigma^{qp}_{eh}$ in the superfluid phase.
The quasiparticle transconductivity $\sigma^{qp}_{eh}$ vanishes as $T \to
T_c$ because, in our theory,  we include no correlation between the two
layers other than the one  implied by the existence of the excitonic
condensate.

Our microscopic calculations are based on the BCS mean field theory
of the transition to the  excitonic superfluid.
Actually, for two-dimensional layers, this transition is expected to be
of the Kosterlitz-Thouless type.   As a result our theory may
require quantitative correction in the region near the transition
temperature where fluctuations are important.  At lower temperatures
our macroscopic analysis shows that the qualitative
behavior of the transresistance is completely
determined by the requirement of least dissipation and is independent of all
microscopic details.  Transresistance measurements should provide
a foolproof test for the presence of an excitonic condensate.
In closing we note that the analysis presented here applies only to the
linear regime; we expect strong non-linearities at temperatures near
the critical temperature.

This work was supported by the NSF  Grants No. DMR-9416906  and DMR-9403908. GV
acknowledges the kind hospitality of the Condensed Matter Theory Group
at Indiana
University,  where this work was initiated. We also acknowledge useful
discussions with Leo Radzihovsky and L. Swierkowski.

\begin{figure}
\vspace{2.5cm}
\hspace{-1cm}
\centerline{\psfig{figure=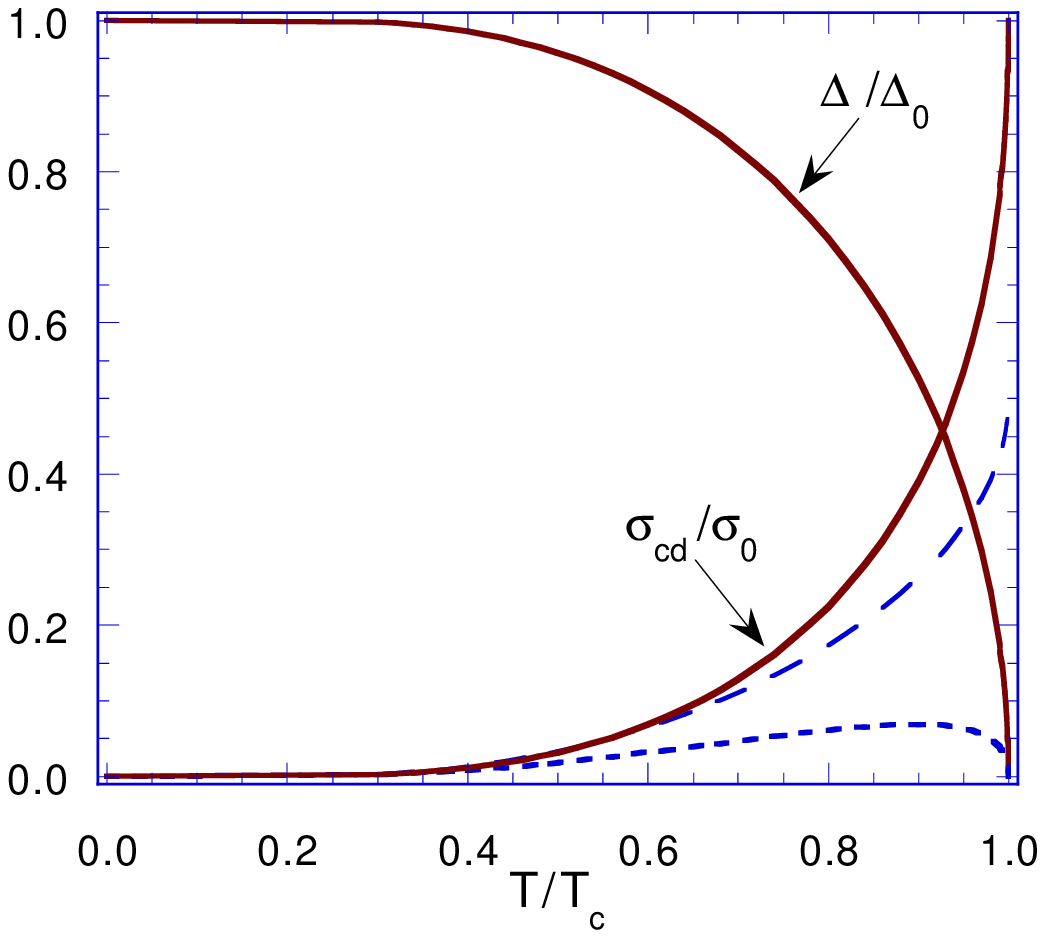,height=7cm,width=7.5cm}}
\vspace{-1.2cm}
\caption{Ratio of the BCS model  condensate drag conductivity
$\sigma_{cd} = \sigma^{qp}_{ee} + \sigma^{qp}_{hh} + 2
\sigma^{qp}_{eh} = \rho_{cd}^{-1}$ to the normal double layer
conductivity $\sigma_0$   as a function of $T/T_c$ for $m_e = m_h$ and
$\tau_{ne} = \tau_{nh}$. Also plotted: quasiparticle conductivities
$\sigma^{qp}_{ee}/\sigma_0$ (long-dashed line), and
$-\sigma^{qp}_{eh}/\sigma_0$ (short dashed line), and   BCS gap
$\Delta$ in units of its zero temperature value $\Delta_0$. }
\label{figure1}
\end{figure}


\begin{references}

\bibitem{oldrussian} S.I. Shevchenko, Fiz. Nizk. Temp. {\bf 2}, 505 (1976)
[Sov. J. Low Temp. Phys. {\bf 2}, 251 (1976)]; Yu. E. Lozovik and
V.I. Yudson, Pis'ma Zh. Eksp. Teor. Fiz. {\bf 22} 556 (1975)
[JETP Lett. {\bf 22}, 271 (1975)]; {\it ibid}, Zh. Eksp. Teor. Fiz.
{\bf 71}, 738 (1976) [Sov. Phys. JETP {\bf 44}, 389 (1976)].

\bibitem{fukuzawa} T. Fukuzawa, E.E. Mendez, and J.M. Hong,
Phys. Rev. Lett. {\bf 64}, 3066 (1990).

\bibitem{kash} J.A. Kash, M. Zachau, E. Mendez, J.M. Hong,
and T. Fukuzawa, Phys. Rev. Lett. {\bf 66}, 2247 (1991).

\bibitem{sivan} U. Sivan, P.M. Solomon, and H. Shtrikman,
Phys. Rev. Lett. {\bf 68}, 1196 (1992).

\bibitem{kane} B.E. Kane, J.P. Eisenstein, W. Wegscheider, L.N. Pfeiffer,
and K.W. West, Appl. Phys. Lett. {\bf 65}, 3266 (1994).

\bibitem{sfcond} Y. Kuramoto and C. Horie, Solid State Commun.
{\bf 25}, 713 (1979); I.V. Lerner, and Yu. E. Lozovik,
Zh. Eksp. Teor. Fiz. {\bf 80}, 1488 (1981) [Sov. Phys. JETP {\bf 53}, 763
(1981)]; Y.A. Bychkov and E.I. Rashba, Solid State Commun. {\bf 48}, 399
(1983); D. Paquet, T.M. Rice, and K. Ueda, Phys. Rev. B {\bf 32}, 5208
(1985).

\bibitem{dlcoh} At strong fields there is an exact mapping between
electron-electron and electron-hole double-layer systems: A.H.
MacDonald and E.H. Rezayi, Phys. Rev. B {\bf 42}, 3224 (1990).
The excitonic condensate state of the electron-hole system
maps to a state with $XY$ pseudospin ferromagnetism in the
electron-electron system.  The Kosterlitz-Thouless transition
temperature for this phase
has been estimated to be as large as $\sim 0.5 K$:
K. Moon {\it et al.}, Phys. Rev. B {\bf 51}, 5138 (1995).  For a
review of broken symmetries of electron-electron double-layer systems
in strong magnetic fields see S.M. Girvin and A.H. MacDonald in
{\it Novel Quantum Liquids in Low-Dimensional Semiconductor Structures},
edited by S. Das Sarma and Aron Pinczuk (Wiley, New York, 1995).

\bibitem{dragold} M.B. Pogrebinskii, Fiz. Tekh. Poluprovodn. {\bf 11},
637 (1977) [Sov. Phys. Semicond. {\bf 11}, 372 (1977)]; Peter J. Price,
Physica {\bf 117 \& 118B}, 750 (1983).

\bibitem{recenttheory} H.C. Tso, P. Vasilopolous, and F.M. Peeters,
Phys. Rev. Lett. {\bf 68}, 1196 (1992); A.-P. Jauho and
H. Smith, Phys. Rev. B {\bf 47}, 4420 (1993); Lian Zheng and A.H. MacDonald,
Phys. Rev. B {\bf 48}, 8203 (1993); Karsten Flensberg and Ben Yu-Kuang, Hu,
Phys. Rev. Lett. {\bf 73}, 3572 (1994); Karsten Flensberg, B. Y. -K Hu,
A. -P. Jahuo, and J. M. Kinaret, preprint (cond-mat 950409-2) 1995;
Alex Kamenev and Yuval Oreg, Phys. Rev. B {\bf 52}, 7516 (1995);
Martin Bonsager and A.-P. Jauho, preprint (1995); L. Swierkowski, J.
Szymanski, and Z. W. Gortel, Phys Rev. Lett {\bf 74}, 3245 (1995); E.
Shimshoni and S. L. Sondhi, Phys. Rev. B {\bf 49} 11484 (1994).

\bibitem{dragexpts} T.J. Gramila, J. P. Eisenstein, A.H. MacDonald,
L.N. Pfeiffer, and K.W. West, Phys. Rev. Lett. {\bf 66}, 1216 (1991);
T.J. Gramila, J.P. Eisenstein, A.H. MacDonald, L.N. Pfeiffer,
and K.W. West, {\sl Surface Sci.}, {\bf 263}, 446 (1992);
T.J. Gramila, J.P. Eisenstein, A.H. MacDonald, L.N. Pfeiffer,
and K.W. West, Phys. Rev. B {\bf 47}, 12957 (1993). See also
P.M. Solomon, P.J. Price, D.J. Frank, and D.C. La Tulipe,
Phys. Rev. Lett. {\bf 63}, 2508 (1989).

\bibitem{whodowecite} S. R. DeGroot and P. Mazur
{\it Non-equilibrium thermodynamics}, (Dover Publications, Inc. New York
1962);  Chapt. 5; I. Prigogine, {\it Introduction to
Thermodynamics of Irreversible Processes}  (Wiley, New York, 1961).

 \bibitem{tinkham} See for example Michael Tinkham, {\it
Introduction to Superconductivity}, (McGraw-Hill, New York, 1975); P.G.
DeGennes
{\it Superconductivity of Metals and Alloys}, (Benjamin, New York,
1966).

\bibitem{caveatbcs} In our calculation we assume that BCS
theory is approximately valid for electron-hole systems but
treat the effective interation which enters the theory, which
is difficult to determine reliably, as a phenomenological parameter.

\bibitem{mahan}G. D. Mahan, {\it Many-particle physics},  (Plenum
Press, New York 1990); Chapter 9.
\end{references}
\end{document}